\documentclass[aps,preprint,nofootinbib,floatfix]{revtex4}
\usepackage{amssymb}
\usepackage{graphicx}
\usepackage{amsmath}

\graphicspath{{./img/}}

\input{tcilatex}

\begin{document}

\title{Spontaneously broken symmetry in string theory}
\author{Jen-Chi Lee }
\email{jcclee@cc.nctu.edu.tw}
\affiliation{Institute of Physics, National Chiao-Tung University, Hsin Chu, Taiwan, ROC}
\date{\today}

\begin{abstract}
By using zero-norm states in the spectrum, we explicitly demonstrate the
existence of an infinite number of high energy symmetry structures of the
closed bosonic string theory. Each symmetry transformation (except those
generated by massless zero-norm states) relates $infinite$ particles with $%
different$ masses, thus they are broken $spontaneously$ at the Planck scale
as previously conjectured by Gross and Evans and Ovrut. As an application,
the results of Das and Sathiapalan which claim that $\sigma $-midel is
nonperturbatively nonrenormalizable are reproduced from a stringy symmetry
argument point of view.
\end{abstract}

\maketitle

%
%-----------------------------------------------------------------------------

%-----------------------------------------------------------------------------

\section{Introduction}

Analogous to the statement that quantum field theory is a quantum mechanical
system with an infinite number degrees of freedom, string theory can be
regarded as an infinite generalization of local quantum field theory with
consistently self-organized couplings. In going from quantum mechanics to
quantum field theory, one suffers from all kinds of high- energy divergences
in the perturbation calculation. But instead, in string theory, one removes
these unwanted divergences by building in an infinite number of high-energy
symmetry structures \cite{1} miraculously when considering the quantum
theory of a free string. In fact, there exist many nonrenormalization
theorems which have been proved up to string two-loop order \cite{2}. It is
believed that this remarkable property of string is due to the existence of
these infinite symmetry structures of the theory. Thus, from a theoretical
point of view, the study of the nonperturbative, high-energy ($\alpha
^{^{\prime }}\rightarrow \infty $), stringy regime of the theory is as
important as, recently developed, 2d quantum gravity \cite{3}, which
promises to extract nonperturbative ( strong-coupling regime) information of
the string. One hopes that the understanding of both nonperturbative regimes 
\cite{4} of the theory may help us to uncover the \textquotedblright
unbroken phase\textquotedblright\ of string theory and shed light on
determining its true vacuum.

\bigskip Gross has shown \cite{1} that there exist an infinite number of
linear relations between the scattering amplitudes of different string
states as $\alpha ^{\prime }\rightarrow \infty $. He then conjectures that
an infinite-parameter symmetry group which is broken spontaneously at the
Plank scale gets restored at very high energy, or $M_{Planck}^{2}\sim
1/\alpha ^{\prime }\rightarrow 0$. On the other hand, it was well known that
the $\sigma $-model can be used to study the dynamics of massless string
modes \cite{5}. This has also been generalized to include higher massive
modes \cite{6}-\cite{8}. Based on the formalism of Evans and Ovrut in Ref. 
\cite{7}, it was proposed \cite{8} that by requiring the decoupling of both
types of zero-norm states in the spectrum, one can derive the complete gauge
symmetries at each fixed mass level. The usual massless Yang-Mills gauge
symmetry and Einstein general covariance can also be generated in this way.
It was remarkable to discover that many higher symmetry transformations
relate particles with different \textquotedblleft spins\textquotedblright\ 
\cite{8}. In this formalism, the dimension of the \textquotedblleft symmetry
group\textquotedblright\ is directly related to the (infinite) number of
zero-norm states in the spectrum. Instead of using the usual $\sigma $-model
loop (or $\alpha ^{\prime }$) expansion \cite{5}, it turns out that the weak
background field approximation (WFA) \cite{8},\cite{9}, valid to all orders
in $\alpha ^{\prime }$, is the appropriate approximation scheme to study the
high-energy symmetry of the string. However, the calculation was done only
in the lowest order WFA. To this order of approximation, one cannot see the
transformation of background fields between different mass sectors, and
hence the $spontaneously$ broken symmetries. The difficulty of higher order
calculation which involves the operator product of two background fields is
closely related to the nonperturbative ( all orders in $\alpha ^{\prime }$,
hence corresponding to the high-energy regime ) nonrenormalizability of the
two-dimensional $\sigma $-model, which has been shown by Das and Sathiapalan 
\cite{10}, and one is forced to introduce counterterms which consist of an
infinite number of massive tensor fields into the theory. In this letter, we
will explicitly demonstrate an infinite number of symmetry transformations
between infinite background fields of $different$ $mass$ $sectors$ of the
closed bosonic string theory. Specifically, we find that for each zero-norm
state whose vertex operator can be written as a worldsheet total derivative,
one can construct a symmetry generator which generates a symmetry
transformation relating an infinite number of particles with different
masses. Hence, together with our previous results in Ref.\cite{8}, where
string states at each fixed mass level form a symmetry multiplet was proved,
we conclude that all string states are connected as an infinite multiplet.
As an interesting application, we also reproduce the results of Das and
Sathiapalan from a stringy symmetry argument point of view.

In the generalized $\sigma $-model formalism, let $T_{\Phi }$ define a
conformal field theory (CFT) with the most general background field
couplings consistent with the string vertex operator consideration in the
WFA ( $\alpha ^{\prime }=1$ ),

\bigskip 
\begin{eqnarray}
T_{\Phi } &=&-\frac{1}{2}\eta _{\mu \nu }\partial X^{\mu }\,\partial X^{\nu
}+h_{\mu \nu }\partial X^{\mu }\,\overline{\partial }X^{\nu }+M_{\mu \nu
,\alpha \beta }\partial X^{\mu }\,\partial X^{\nu }\overline{\partial }%
X^{\alpha }\overline{\partial }X^{\beta }+D_{\mu \nu ,\alpha }\partial
X^{\mu }\,\partial X^{\nu }\overline{\partial ^{2}}X^{\alpha }  \notag \\
&&+E_{\mu ,\alpha \beta }\partial ^{2}X^{\mu }\,\overline{\partial }%
X^{\alpha }\overline{\partial }X^{\beta }+A_{\mu ,\alpha }\partial
^{2}X^{\mu }\overline{\partial ^{2}}X^{\alpha }+...+\text{ }higher\text{ }%
order\text{ }terms,  \label{1}
\end{eqnarray}%
with background fields ($\Phi $) equations of motion

\bigskip 
\begin{equation}
\beta _{i}[\Phi ]=0  \label{2}
\end{equation}%
where $\beta _{i}$ are the renormalization group $\beta $ functions for each
background coupling. We have used the worldsheet light-cone coordinates in (%
\ref{1}) and neglected a similar left-moving equation \cite{7,8}. In the
first order WFA, we have calculated many examples which involve lower
massive states \cite{8}. On the other hand, it has been demonstrated \cite{7}
that if one can find an infinitesimal operator Q such that \bigskip 
\begin{equation}
T_{\Phi }+[Q,T_{\Phi }]=T_{\Phi +\delta \Phi },  \label{3}
\end{equation}%
then the worldsheet generator Q generates a space-time symmetry
transformation. In that case, $T_{\Phi +\delta \Phi }$ specifies a new CFT
with $\beta _{i}[\Phi +\delta \Phi ]=0.$ In the first order WFA, it can be
shown that the integral of each worldsheet (1,0) or (0,1) primary field
corresponds to a Q which fulfils the criterion in (\ref{3}). A key step,
made in Ref.\cite{8}, was to realize that the complete gauge symmetries of
the string which include those in (\ref{3}) can be systematically
constructed by using the well-known zero-norm states in the spectrum. Hence,
as one expects for a unitary theory, all space-time symmetries are directly
related to the decoupling of zero-norm states in the spectrum. In this
paper, however, we will calculate the important higher order correction of
the symmetry transformations. It is from this second order correction (
first order in the background fields and first order in the transformation
parameters ) that one begins to see the nonperturbative character (
high-energy character ) of the symmetry transformations, and hence the
spontaneously broken symmetries. We will use Eq.(\ref{3}) to do the
calculation for Q constructed by those zero-norm states whose vertex
operators can be written as a worldsheet total derivative. As an example, we
give a class of type I zero-norm states ( states with zero norm in any
space-time dimension) \cite{8} of the following form ( omit all spin indices
):

\begin{equation}
\theta (\alpha _{-1})^{n+1}(\widetilde{\alpha }_{-1})^{n-1}\widetilde{\alpha 
}_{-2}|0,k>+k\theta (\alpha _{-1})^{n+1}(\widetilde{\alpha }%
_{-1})^{n+1}|0,k>,  \label{4}
\end{equation}%
where $\theta $ is a $2n+1$ spin index parameter that is symmetric on the
first $n+1$ and last $n$ indices. It is orthogonal to $k_{\mu }$ on each
index, and traceless on any pair of the first $n+1$, or any pair of the last 
$n$ indices. In the second term, the index of $k_{\mu }$ is symmetrized with
the last $n$ indices of $\theta $. The mass of the state is$\bigskip $%
\begin{equation}
-k^{2}=M^{2}=2n.  \label{5}
\end{equation}%
The corresponding worldsheet generator is%
\begin{equation}
Q_{n}=\int d\sigma \theta (X)(\partial X)^{n+1}(\partial \overline{X})^{n},
\label{6}
\end{equation}%
where $\theta $ \ is now promoted to become a function of X, and the
orthogonality condition stated above becomes divergence free on each index.
Also, Eq.(\ref{5}) is replaced by $(\Box -2n)\theta =0$. Under these
constrains, it can be shown that the integrand in Eq.(\ref{6}) is a (1,0)
primary field. The lowest order ( zero order in the background fields and
first order in the $\theta $ parameters ) calculation of $[Q_{n},T_{\Phi }]$
gives\bigskip 
\begin{equation}
\partial _{\nu }\theta (X)(\partial X)^{n+1}(\overline{\partial }%
X)^{n+1}+\theta (X)(\partial X)^{n+1}(\partial ^{2}X)(\overline{\partial }%
X)^{n-1},  \label{7}
\end{equation}%
where the index $\nu $ is symmetrized with the last $n$ indices of $\theta $%
. Comparing with Eq.(\ref{1}) and using the constraints on $\theta $, we
find that Eg.(\ref{7}) generates a symmetry transformation for a single $n$%
th massive level particle. The nonperturbative effects begin to show up at
the second order calculation. In the following, we will use the lowest order
results to calculate the second order correction of the symmetry
transformation law. In general, one has to calculate terms of the following
type ( we use complex coordinates in this calculation ), 
\begin{equation}
\lbrack Q_{n},M(X)(\partial X)^{m+1}(\overline{\partial }X)^{m+1}]=\int 
\frac{d\omega }{2\pi i}\theta (X)(\partial X)^{n+1}(\overline{\partial }%
X)^{n}(\omega )\cdot M(X)(\partial X)^{m+1}(\overline{\partial }X)^{m+1}(z),
\label{8}
\end{equation}%
and compare them with the first order terms of the background fields in Eq.(%
\ref{1}) to see whether they satisfy Eq.(\ref{3}) or not. It can be checked 
\cite{11} that the only dangerous terms which might violate Eq.(\ref{3})
consist of $operator$ $contraction$ of the form $<\theta (X(\omega
))M(X(z))> $ in the integrand. To prove that those terms vanish, let 
\begin{equation}
\theta (X)=\int dk\theta (k)e^{ikx},\text{ }M(X^{\prime })=\int dk^{\prime
}M(k^{\prime })e^{ik^{\prime }x^{\prime }},  \label{9}
\end{equation}%
where $k=(k_{0},k)$ is the 26d momentum. From the lowest order calculation
[8]. we have%
\begin{equation}
(\Box -2n)\theta =0,\text{ }(\Box -2m)M=0,  \label{10}
\end{equation}%
which means

\begin{equation}
k^{2}=-2n,\text{ }k^{\prime 2}=-2m,  \label{11}
\end{equation}%
Eg.(\ref{11}) looks like on-shell conditions although we are not calculating
scattering amplitude. They are valid only in the lowest order calculation.
So, for each fixed $n\geq 1$, one has to deal with integral of the following
form $(s\leq Min[n+1,m+1])$:%
\begin{eqnarray}
I &=&\int dkdk^{\prime }\theta (k)M(k^{\prime })\oint \frac{d\omega }{2\pi i}%
<e^{ikx(\omega )}e^{ik^{\prime }x(z)}>  \notag \\
\times &<&\partial X(\omega )\partial X(z)>^{s}(\partial X(\omega ))^{n+1-s}(%
\overline{\partial }X(\omega ))^{n}(\partial X(z))^{m+1-s}(\overline{%
\partial }X(z))^{m+1},  \label{12}
\end{eqnarray}%
where $m$ is any nonnegative integer. In the background field method, $%
<e^{ikx(\omega )}e^{ik^{\prime }x(z)}>$ is defined to be%
\begin{equation}
<e^{ikx(\omega )}e^{ik^{\prime }x(z)}>=\int [d\xi ]e^{ik(x_{0}(\omega )+\xi
(\omega ))}e^{ik^{\prime }(x_{0}(z)+\xi (z))}e^{-S[x_{0}+\xi ]-S[x_{0}]},
\label{13}
\end{equation}%
where $X^{\mu }=X_{0}^{\mu }+\xi ^{\mu }$ is expanded around a classical
background $X^{\mu }$ and $\xi ^{\mu }$ is the quantum fluctuation. The
worldsheet action is%
\begin{equation}
S=\int d^{2}z\partial X^{\mu }\overline{\partial }X^{\nu }\eta _{\mu \nu }.
\label{14}
\end{equation}%
The calculation of (\ref{13}) is straightforward, one gets%
\begin{equation}
e^{ikx_{0}(\omega )}e^{ik^{\prime }x_{0}(z)}\mu ^{k^{2}+k^{\prime 2}}(\mu
\left\vert \omega -z\right\vert )^{2kk^{\prime }},  \label{15}
\end{equation}%
where $\mu $ is an infrared cutoff. The factor $\mu ^{k^{2}+k^{\prime 2}}$
comes from the tadpole divergences which occurred already in the V.E.V. of
single vertex function. To cancel the infrared cutoff dependence $\mu $, we
must require $k+k^{\prime }=0$ or Eq.(\ref{15}) will vanish as $\mu $ goes
to zero. Hence%
\begin{eqnarray}
&<&e^{ikx(\omega )}e^{ik^{\prime }x(z)}>=e^{ikx_{0}(\omega )}e^{ik^{\prime
}x_{0}(z)}\left\vert \omega -z\right\vert ^{-k^{2}-k^{\prime 2}}\text{ for }%
k+k^{\prime }=0,\text{\ }  \TCItag{16a} \\
\text{\ } &<&e^{ikx(\omega )}e^{ik^{\prime }x(z)}>=0\text{ \ \ \ \ \ \ \ \ \
\ \ \ \ \ \ \ \ \ \ \ \ \ \ \ \ \ \ \ \ \ \ \ \ \ \ \ \ \ for }k+k^{\prime
}\neq 0.  \TCItag{16b}
\end{eqnarray}%
By using Eq.(\ref{11}) which is the result of the lowest order calculation,
we note that Eq.(\ref{16a}a) contributes a factor $\left\vert \omega
-z\right\vert ^{2(n+m)}=\left\vert \omega -z\right\vert ^{4n}$ in the
integrand of $I$. But $<\partial X(\omega )\partial X(z)>^{s}$contributes $%
\left\vert \omega -z\right\vert ^{-2S}$ to $I$. Since $s\leq n+1$, we
conclude that $I$ vanishes for $n\geq 1$. It is important to note that Eq.(%
\ref{11}) is crucial to prove our final result, or $I$ can be divergent for
some range of $(k,k^{\prime })$. To be concrete, we give the $n=1$ case as
an example which corresponds to the worldsheet generator 
\begin{equation}
Q_{i}=\oint \frac{d\omega }{2\pi i}\theta _{\mu \nu ,\alpha }\partial X^{\mu
}\partial X^{\nu }\overline{\partial }X^{\alpha }.  \tag{17}  \label{17}
\end{equation}%
The calculation of $[Q_{i},h_{\mu ,\nu }\partial X^{\mu }\partial X^{\nu
}+...]$ with all first order background fields included is straightforward.
One gets the following infinite symmetry transformation:%
\begin{eqnarray*}
\delta M_{(\alpha \beta ),\lambda \nu } &=&\partial _{\nu }\theta _{\alpha
\beta ,\lambda }-2\partial _{\beta }\theta _{\alpha ,\lambda }^{\gamma
}h_{\gamma ,\nu }-2\theta _{\alpha ,\lambda }^{\gamma }\partial _{\gamma
}h_{\beta ,\nu }-\partial ^{\nu }\theta _{\alpha \beta ,\lambda }h_{\mu ,\nu
}+\partial _{\alpha }\partial _{\beta }\theta _{,\lambda }^{\gamma \mu
}\partial _{\gamma }h_{\mu ,\nu } \\
&&+\partial _{\alpha }\theta _{,\lambda }^{\gamma \delta }\partial _{\gamma
}\partial _{\delta }h_{\beta ,\nu }-2\partial _{\alpha }\partial ^{\mu
}\theta _{\beta ,\lambda }^{\delta }\partial _{\delta }h_{\mu ,\nu }+\frac{1%
}{2}\partial _{\alpha }\partial _{\beta }\partial ^{\mu }\theta _{,\lambda
}^{\gamma \delta }\partial _{\gamma }\partial _{\delta }h_{\mu ,\nu },
\end{eqnarray*}%
\begin{equation*}
\delta D_{(\alpha \beta ),\lambda }=\theta _{\alpha \beta ,\lambda },
\end{equation*}%
\begin{equation*}
\delta E_{\alpha ,(\lambda \nu )}=-2\theta _{\alpha ,\lambda }^{\beta
}h_{\beta ,\nu }+\partial _{\alpha }\theta _{,\lambda }^{\gamma \mu
}\partial _{\gamma }h_{\mu ,\nu }-2\partial ^{\mu }\theta _{\alpha ,\lambda
}^{\gamma }\partial _{\gamma }h_{\mu ,\nu }+\frac{1}{2}\partial _{\alpha
}\partial ^{\mu }\theta _{,\lambda }^{\gamma \delta }\partial _{\gamma
}\partial _{\delta }h_{\mu ,\nu },
\end{equation*}%
\begin{equation*}
\delta A_{\alpha ,\beta }=0,
\end{equation*}%
\begin{equation}
...,  \tag{18}  \label{18}
\end{equation}%
where the zero order ( in background fields ) terms have been calculated in
Eq.(\ref{7}). The symmetric property of the spin indices on the r.h.s. of
Eq.(\ref{18}) is understood. One can calculate the transformations
corresponding to all higher massive modes as well \cite{11}. Each single
transformation in Eq.(\ref{18}) relates particles with mass difference one.
Thus, this $n=1$ massive zero-norm state can be used to generate a symmetry
transformation which relates all particles in the bosonic string spectrum (
except tachyon ). Similar argument goes for general $Q_{n}$ cases. The
symmetry generated by $Q_{n}$ relates particles with mass difference $n$ .
For $n=0$ case, $I=0$ if $m=0$. In that case, one can still check that
contribution of $I$ does not violate Eq.(\ref{3}) . Indeed, an explicit
calculation gives%
\begin{eqnarray}
\delta h_{\mu ,\nu } &=&\partial _{\nu }\theta _{\mu }+\partial ^{\alpha
}\theta _{\mu }h_{\alpha ,\nu }-2\partial _{\mu }\theta ^{\alpha }h_{\alpha
,\nu }-\theta ^{\lambda }\partial _{\lambda }h_{\mu ,\nu }-\partial _{\mu
}\partial ^{\alpha }\theta ^{\lambda }\partial _{\lambda }h_{\alpha ,\nu }, 
\notag \\
&&...\text{ }.  \TCItag{19}  \label{19}
\end{eqnarray}%
One can also calculate the transformations of all higher massive modes \cite%
{11}. Note that, to any finite order calculation in WFA, the usual general
covariance for graviton is lost. The symmetry transformation corresponding
to $Q_{0}$ is the only symmetry which relates particles with the same mass.
Therefore, one is tempted to argue that this infinite parameter
\textquotedblleft symmetry group\textquotedblright constructed from $Q_{n}$
is broken spontaneously down to $Q_{0},$ leaving the corresponding gauge
particle massless. Presumably the Higgs mechanism is operating as suggested
by the inhomogeneous terms of Eq.(\ref{18}) . All Goldstone bosons become
the longitudinal parts of higher massive modes in the string spectrum.

There is another interesting implication of the present work. If one
believes that the symmetry structures presented in this paper are crucial in
the study of the quantum theory of string, then any truncation of massive
modes would inevitably lose these important symmetry structures, and thus
leads to meaningless results. This is just like the case of Kaluza-Klein
truncation \cite{12}. In fact, when one proved the nonrenormalization
theorems for massless external legs, the massive modes effects have been
included in the string-loop diagram. Unfortunately, until now, most
researches of the theory of string have been confined to the low energy
regime. The results of this paper strongly suggest that nonperturbative
string vacuum should be seriously considered. We believe that there are
still many fundamental structures in the high-energy regime which remain to
be uncovered even in the critical string theory. An interesting application
of the symmetry presented in Eq.(\ref{18}) is following: \ If one naively
includes only the massless mode in Eq.(\ref{1}), then the symmetry induced
by $Q_{1}$ in Eq.(\ref{18}) will force one to include all higher massive
modes. This simple observation is consistent with the results of Das and
Sathiapalan \cite{10} and the fact that there are infinite couplings between
infinite number of states of the string \cite{9,13}. Finally, there are
still many zero-norm states which cannot be written as a worldsheet total
derivative \cite{8}. Further studies are in progress.

From the BRST point of view, the $zeroth-order$ $WFA$ of our approach is
equivalent to the fact that the states $Q_{BRST}|\Psi >$ with definite ghost
number are zero-norm states and should be decoupled from the physical
S-matrix. This is analogous to the BRST formulation of usual Yang-Mills
theory where we know exactly a priory what the classical action is from the
symmetry principle. It is thus easy to convince oneself that string theory
can be regarded as a spontaneously broken Yang-Mills type theory with
infinite dimensional \textquotedblleft gauge group\textquotedblright
constructed by an infinite number of zero-norm states in the spectrum.

\bigskip

I would like to thank Prof. S. C. Lee for inviting me to IPAS where this
work was finished. I am grateful also to Profs. B. A. Ovrut and M. Evans for
useful suggestion in the early stage of this work, and Profs. C.R. Lee and
W.K.Sze for stimulating discussion. This work was supported in part by the
National Science Council of the R.O.C. under contract Nos.
NSC80-0208-M009-34 and NSC81-0115-C001-03.

%-----------------------------------------------------------------------------
%\bibliographystyle{amsplain}

%-----------------------------------------------------------------------------

\end{document}